\documentclass[a4paper]{jpconf}
\usepackage{graphicx}

\usepackage{epsfig}
\usepackage{amsmath}
\usepackage{amsfonts}
\usepackage{amssymb}
\usepackage{url}
\usepackage{hyperref}
\usepackage{subfigure}
\usepackage{cancel}
\newcommand{\be}{\begin{equation}}
\newcommand{\ee}{\end{equation}}
\newcommand{\bea}{\begin{eqnarray}}
\newcommand{\eea}{\end{eqnarray}}
\newcommand{\nn}{\nonumber}

\begin{document}
\title{Implications of lepton nonuniversality in the beauty sector}
\author{A. Giri$^{a*}$, R. Mohanta$^b$ and S. Sahoo$^b$}
\address{$^a$Physics Department,   IIT Hyderabad, Kandi - 502285, India\\
$^b$School of Physics, University of Hyderabad, Hyderabad - 500046, India }
\ead{$^*$~ giria@iith.ac.in}

\begin{abstract} 

The phenomenon of CP violation in the standard model (SM) framework and the decay dynamics have been established from the data obtained from the B factories and so far we have not seen anything new. Nevertheless, there have been instances of deviations in many measured observables in the flavor sector, as far as the data and predictions are concerned. Here we will mention some deviations obtained in measurements related to lepton universality, as seen from the data, and try to understand their implications. To accommodate the observed data we will consider a leptoquark model, which seems to be one interesting model beyond the SM. 
\end{abstract}

\section{Introduction}
\vspace*{0.1 truecm}
The course of high energy physics is going through an interesting and exciting phase.  We have in one hand many experiments, conducted in the last few decades, measured various observables which are in excellent agreement with the model proposed by Glashow, Salam and Weinberg, which is also known as the standard model in the literature. One exception being that of the observation of neutrino oscillation. At the same time, there are many fundamental unsolved questions like the hierarchy problem, dark matter and baryon asymmetry of the Universe etc., which make ourselves believe that there is something beyond that of the SM. In fact, the resounding success of the SM has led us with no option but to believe that whatever may be the form of the new physics the low energy limit of the same is the SM.  The $V-A$ current structure of the weak interactions has been established long ago and, in fact, in the past has been very instrumental in providing many interesting and accurate results.  It is interesting to note that we have observed, in the last few years, some kind of unusual results involving leptons, in particular the heavy $\tau$ (or the third generation leptons). The belief is that the couplings of third generation fermions to the electroweak symmetry breaking is comparatively stronger due to their large masses and therefore, sensitive to new physics that modifies the $V-A$ structure of the SM.  From this point of view the study of $B^{(*)} \to \tau \bar{\nu}$ and $B \to D^{(*)} \tau \bar{\nu}$ charge current processes are really interesting. In recent measurements,  BaBar \cite{Babar} and Belle \cite{Belle} have reported $3.5\sigma$ deviation  in the  ratio of branching fractions of $\bar B \to \bar D^{(*)} \tau \bar \nu$  over $\bar B \to \bar D^{(*)} l \bar \nu$, where $l=e, \mu$, 
\begin{equation}
R_{D}=\frac{{\rm Br}\left(\bar{B} \to D \tau \bar \nu \right)}{{\rm Br}\left(\bar{B} \to D l \bar \nu \right)} = 0.421 \pm  0.058, \nonumber \\~~~
R_{D^*}=\frac{{\rm Br}\left(\bar{B} \to D^* \tau \bar \nu \right)}{{\rm Br}\left(\bar{B} \to D^* l \bar \nu \right)} = 0.337 \pm  0.025,
\end{equation}
from their corresponding SM predictions,
\begin{equation}
R_D^{\rm SM} = 0.305 \pm 0.012, ~~~~~~ R_{D^*}^{\rm SM} = 0.252 \pm 0.004.
\end{equation}
The above results might indicate the violation of lepton universality. It should be noted here that the observables measured, as mentioned above, are ratio of two processes, where in the numerator and denominator the initial state is the same, whereas the final sates differ depending upon the kind of leptons involved. This actually helps us to reduce the theoretical uncertainties since most of the contributing terms are same and uncertainties actually cancel out. Since these decays occur at the tree level in the SM, the general expectation is that models with masses of the new particles near the TeV scale are required to explain the anomaly. The branching ratios of semileptonic $b \to c l \bar \nu$ processes  can be computed precisely due to the  light leptons mass, thus the deviation in $R_{D^{(*)}}$  is obviously from the new physics affecting the $\bar B \to \bar D^{(*)} \tau \bar \nu$ process.  The branching ratios and $R_{D^{(*)}}$ anomaly in the SM and in various new physics models have been investigated  in the literature.  Similarly, another interesting observable, reported by LHCb \cite{RK}, is the lepton non universality in  $B \to K l^+ l^-$ process, 
\begin{equation}
R_K=\frac{{\rm Br}\left(\bar{B} \to K \mu^+ \mu^- \right)}{{\rm Br}\left(\bar{B} \to K  e^+ e^- \right)} = 0.745^{+0.090}_{-0.074} \pm 0.036,
\end{equation}
which has $2.6\sigma$ deviation from the SM value, $R_K = 1.0003 \pm 0.0001$, in the dilepton invariant mass squared bin $\left( 1 \leq q^2 \leq 6 \right) {\rm GeV^2}$. In the semileptonic decay, the rate of $B \to K^* \mu^+ \mu^-$ and the angular observables $P_5^\prime$ \cite{p5prime}  have   $3\sigma$ deviations from the SM predictions. The discrepancy of $3\sigma$ is also found in the decay rate of the $B_s \to \phi \mu^+ \mu^-$ process \cite{phi-decayrate}.

In this article, we pursue the analysis of rare semileptonic decays of $B$ meson to leptons of second and third generations and we extend the SM by an additional leptoquark model which is built based on the  $SU(3)_c \times SU(2)_L \times U(1)_Y$ gauge symmetries. The study of famous $R_{D^{(*)}}$ anomaly and the lepton non-universality in the $b \to c \tau \bar{\nu}$ decay processes are the main interests of this work. We calculate the branching ratios, forward-backward asymmetries and the $\tau$-polarisations of $B \to D^{(*)} \tau \bar{\nu}$ processes in the leptoquark model. We estimate the branching ratio of rare leptonic $B_{u, c}^* \to \tau \bar{\nu}$ decay process of $B^*_{u,c}$ vector meson. In the leptoqaurk model, we also explore the possibility of lepton nonuniversality parameters in the $B^* \to \tau \bar{\nu}$  and $\Lambda_b \to \Lambda_c \tau \bar{\nu}$ processes. Leptoquarks can couple  (decay) to a quark and lepton of the same generation simultaneously and carry both lepton and baryon number. They can have spin $0$ (scalar leptoquarks) or spin $1$ (vector leptoquarks) and can be characterized by their fractional electric  charge and Fermion numbers $F=3B+L$, where $B$ and $L$ are the baryon number and lepton number respectively. Such leptoquarks exist in some extended SM theories \cite{georgi} such as grand unified theories based on $SU(5)$, $SU(10)$, Pati-Salam model, technicolor model  and composite model etc. To avoid rapid proton decay, we consider that the leptoquark does not couples to diquarks and therefore conserve baryon and lepton numbers. The leptoquark model in the context of $B$-physics anomalies has been taken up in the literature.

The outline of this paper is follows. In section 2, we describe the effective Hamiltonian involving $b \to c \tau \bar \nu$ quark level transition  in the SM.  In section 3 we discuss the new physics contributions coming from vector leptoquarks and show how they can explain the observed anomalies in $b$-sector. Our results are presented in Section 4. 

\section{ Effective Hamiltonian for $b \to c \tau \bar{\nu}_l$ and $b \to s l^+ l^-$ processes }
\vspace*{0.1 truecm}
In this section we write the relevant effective Hamiltonian in the SM as given by \cite{sakaki} 

\begin{equation}
\mathcal{H}_{eff}=\frac{4G_F}{\sqrt{2}} V_{cb} \Big [ \left(\delta_{l\tau} + C_{V1}^l \right) \mathcal{O}_{V1}^l + C_{V2}^l \mathcal{O}_{V2}^l +  C_{S1}^l \mathcal{O}_{S1}^l +  C_{S2}^l \mathcal{O}_{S2}^l+ C_{T}^l \mathcal{O}_{T}^l \Big ],
\end{equation}
where $G_F$ is the Fermi constant, $V_{cb}$ is the Cabibbo-Kobayashi-Maskawa (CKM) matrix element and the index $l$ stands for neutrino flavour, $l=e, \mu, \tau$. The $C_X^l$'s, where $X=V_{1,2}, S_{1,2}, T$ are the Wilson coefficients and the corresponding operators are
\begin{eqnarray}
&&\mathcal{O}_{V1}^l = \left(\bar{C}_L \gamma^\mu b_L \right) \left(\bar{\tau}_L \gamma_\mu \nu_{lL} \right),\nn\\
&&\mathcal{O}_{V2}^l = \left(\bar{C}_R \gamma^\mu b_R \right) \left(\bar{\tau}_L \gamma_\mu \nu_{lL} \right),\nn\\
&&\mathcal{O}_{S1}^l = \left(\bar{C}_L  b_R \right) \left(\bar{\tau}_R \nu_{l L} \right),\nn\\
&&\mathcal{O}_{S2}^l = \left(\bar{C}_R b_L \right) \left(\bar{\tau}_R \nu_{l L} \right),\nn\\
&&\mathcal{O}_{T}^l = \left(\bar{C}_R \sigma^{\mu \nu}  b_L \right) \left(\bar{\tau}_R \sigma_{\mu \nu} \nu_{lL} \right),
\end{eqnarray}
where $L(R)=(1\mp \gamma_5)/2$ are the projection operators. Since the flavour of neutrino is not observed at $B$-factories, all generations of neutrinos can be taken into the account to reveal the signature of new physics. In the SM, the contribution to the $b \to c l \bar{\nu}_l$ process is indicated as $\delta_{l \tau}$ and  the Wilson coefficients ($C_X^l$) are  zero, which can only be generated by new physics models. These new couplings can be bound experimentally, so that the effects of the new operators can be scrutinized in physical observables.

Similarly, the effective Hamiltonian describing the processes induced by $b \to s l^+ l^-$ transitions in the SM is given by  \cite{b-s-Hamiltonian}
\bea
{\cal H}_{eff} &=& - \frac{ 4 G_F}{\sqrt 2} V_{tb} V_{ts}^* \Bigg[\sum_{i=1}^6 C_i(\mu) O_i +C_7 \frac{e}{16 \pi^2} \Big(\bar s \sigma_{\mu \nu}
(m_s P_L + m_b P_R ) b\Big) F^{\mu \nu} \nn\\
&&+C_9^{eff} \frac{\alpha}{4 \pi} (\bar s \gamma^\mu P_L b) \bar l \gamma_\mu l + C_{10} \frac{\alpha}{4 \pi} (\bar s \gamma^\mu P_L b)
\bar l \gamma_\mu \gamma_5 l\Bigg]\;,\label{ham}
\eea
where $\alpha$ is the fine structure constant, $V_{tb}V_{ts}^*$ is the product of CKM matrix element and $C_{i}$'s are the Wilson coefficients evaluated at the renormalization  scale $\mu=m_b$ \cite{b-s-Wilson}.
In the following subsections we will explain the possible leptoquarks relevant for the $b \to c l \bar{\nu}_l$ and $b \to s l^+  l^-$ quark level transitions.

\section{Vector leptoquarks and new Physics }
\vspace*{0.1 truecm}

Here we consider the new physics model in which the new particle interacts both with quarks and leptons simultaneously, called leptoquark, and carries both the baryon and lepton numbers.
Leptoquarks have ten different multiplets \cite{Ruckl}  under the  $SU(3)_c \times SU(2)_L \times U(1)_Y$ SM gauge symmetries with flavour non-diagonal couplings. Half of them have scalar nature and other  halves have vector nature under the Lorentz transformation. The scalar (vector) leptoquarks have spin $0~(1)$ and could potentially contribute to the FCNC processes involving the quark level transitions $b \to s l^+ l^-$ and $b \to c l^- \bar{\nu}$.   Out of all possible leptoquark multiplets, six scalar and vector leptoquark bosons  are relevant for the $b \to c l \bar{\nu}$ processes. Here $S_{1, 3}$ and $R_2$ are the scalar leptoquark bosons, $U_{1, 3}^\mu$ and $V_2^\mu$ are the vector leptoquark bosons. The  vector leptoquarks with charge$=2/3$ and with 0 fermion noumber can mediate both $b \to s l^+ l^-$ and $b \to c l^- \bar{\nu}$ quark level transitions. Therefore, $U_1^\mu=(3, 3, 2/3)$ and $U_3^\mu=(3, 1, 2/3)$ (where the numbers in the parenthesis represent the respective quantum numbers under the standard model gauge group, $SU(3)_c \times SU(2)_L \times U(1)_Y$) are only valid vector leptoquarks to study both $R_{K^{(*)}}$ and $R_{D^{(*)}}$ anomaly. 
In this work, we investigate  the $U_1^\mu=(3, 3, 2/3)$ and $U_3^\mu=(3, 1, 2/3)$ vector leptoquarks, which have charge=2/3, fermion no =0 and can mediate  both   $b \to s l^+ l^-$ and $b \to c l^- \bar{\nu}$ quark level transitions.
In order to avoid rapid proton decay, we do not consider diquark interactions, as the presence of both leptoquark and diquark interactions will violate baryon and lepton number.

The interaction Lagrangian of $U_{1, 3}^\mu$ leptoquarks with the SM fermion bilinear is given as \cite{sakaki, Ruckl}

\begin{equation}
{\cal{L}}^{LQ}=\left(h_{1L}^{ij}\bar{Q}_{iL} \gamma^\mu L_{jL} + h_{1R}^{ij}\bar{d}_{iR} \gamma^\mu l_{jR} \right) U_{1\mu} + h_{3L}^{ij}\bar{Q}_{iL} \pmb{\sigma}  \gamma^\mu L_{jL} U_{3\mu}, \label{Lagrangian}
\end{equation}

where $Q_L (L_L)$ is the left handed quark (lepton) doublet, $u_R (d_R)$ and $l_R$ are the right-handed  up (down) quark and lepton singlet respectively and $\psi^c = C \bar{\psi}^T = C \gamma^0 \psi^*$ is the charge-conjugated fermion field of $\psi$.  The leptoquark couplings are represented by  $h^{ij}$, where $i, j$ are the generation indices of quarks and leptons respectively.

 Here the fermions are stated in the gauge eigen basis in which  Yukawa couplings of the up type quarks and the charged leptons are diagonal, whereas the down-type quark fields are rotated into the mass eigenstate basis by the CKM matrix. Now performing the Fierz transformations, we obtain the additional Wilson coefficients to the $b \to c \tau \bar{\nu}_l$ process as \cite{Ruckl},

\begin{subequations}
\bea
&&C_{V_1}^l=\frac{1}{2\sqrt{2}G_F V_{cb}}\sum_{k=1}^3 V_{k3}\Bigg [ \frac{h_{1L}^{2l}{h_{1L}^{k3}}^*}{2M^2_{U_1^{2/3}}} - \frac{h_{3L}^{2l}{h_{3L}^{k3}}^*}{2M^2_{U_3^{2/3}}} \Bigg ], \label{CV1} \\
&& C_{V_2}^l=0,  \label{CV2} \\
&& C_{S_1}^l = -\frac{1}{2\sqrt{2}G_F V_{cb}}\sum_{k=1}^3 V_{k3} \frac{2 h_{1L}^{2l}{h_{1R}^{k3}}^*}{M^2_{U_1^{2/3}}} ,
\label{CS1}
\eea
\end{subequations}

where $V_{k3}$ denotes the CKM matrix element and $M_{U_{1(3)}^{2/3}}$ is the mass of the leptoquark. 

After expanding the $SU(2)$ indices of Eqn. (\ref{Lagrangian}), one can notice that  $U_{1, 3}$ vector leptoquarks also contributes additional Wilson coefficients to the $b \to s l^+ l^-$ processes as 

\begin{subequations}
\begin{align}
  C_{9}^{NP} &= -C_{10}^{NP}  = \frac{1}{2\sqrt{2} G_F V_{tb}V_{ts}^*}
\Big [  \frac{h_{1L}^{2l}{h_{1L}^{k3}}^*}{M^2_{U_1^{2/3}}} +
 \frac{h_{3L}^{2l}{h_{3L}^{k3}}^*}{M^2_{U_3^{-1/3}}} \Big ] \,.  \\
 C_9^{\prime NP} &= C_{10}^{\prime NP} = \frac{1}{2\sqrt{2} G_F V_{tb}V_{ts}^*}  \frac{h_{1R}^{2l}{h_{1R}^{k3}}^*}{M^2_{U_1^{2/3}}}\,,\\
  -C_P^{NP} &= C_{S}^{NP} = \frac{1}{\sqrt{2} G_F V_{tb}V_{ts}^* }
   \frac{h_{1L}^{2l}{h_{1R}^{k3}}^*}{M^2_{U_1^{2/3}}}\,,\\
  C_P^{\prime NP} &= C_{S}^{\prime  NP}= \frac{1}{\sqrt{2} G_F V_{tb}V_{ts}^* } \frac{h_{1R}^{2l}{h_{1L}^{k3}}^*}{M^2_{U_1^{2/3}}}\,.
\end{align}
\end{subequations}

\section{Results and Discussion}
\vspace*{0.1 truecm}
In order to use the scenario of vector leptoquarks we constrain the parameter space in terms of the couplings from the existing data and thereafter use the same values to explain the anomalies, as mentioned above \cite{mohanta2}. Looking at the figures it may be concluded that the possibility of vector type of leptoquarks can be thought of as an alternative option at least in the context of the subject matter discussed here.

\begin{figure}[h]
\centering
\includegraphics[width=7.5 cm,height=5.5cm]{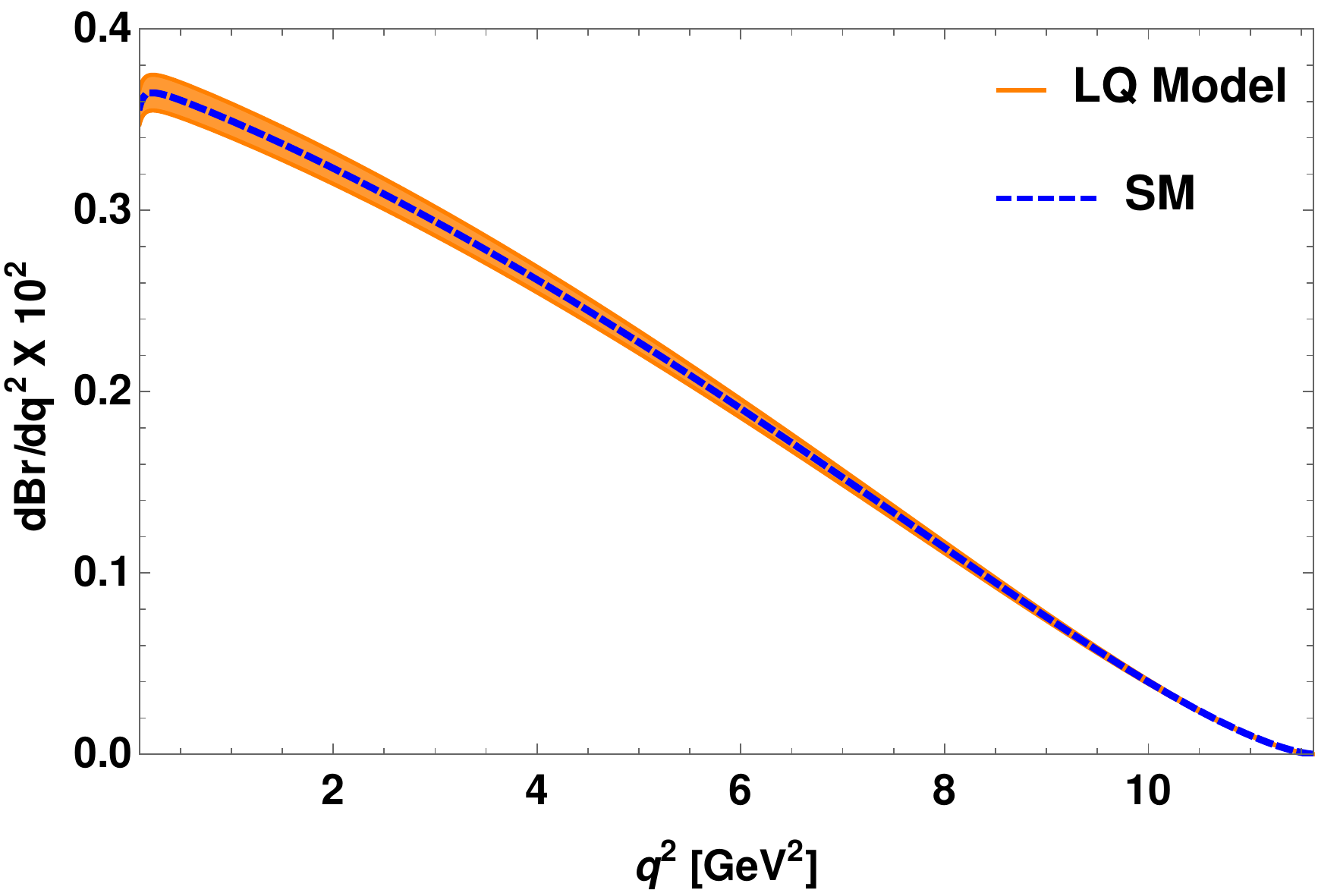}
\quad
\includegraphics[width=7.5 cm,height=5.5cm]{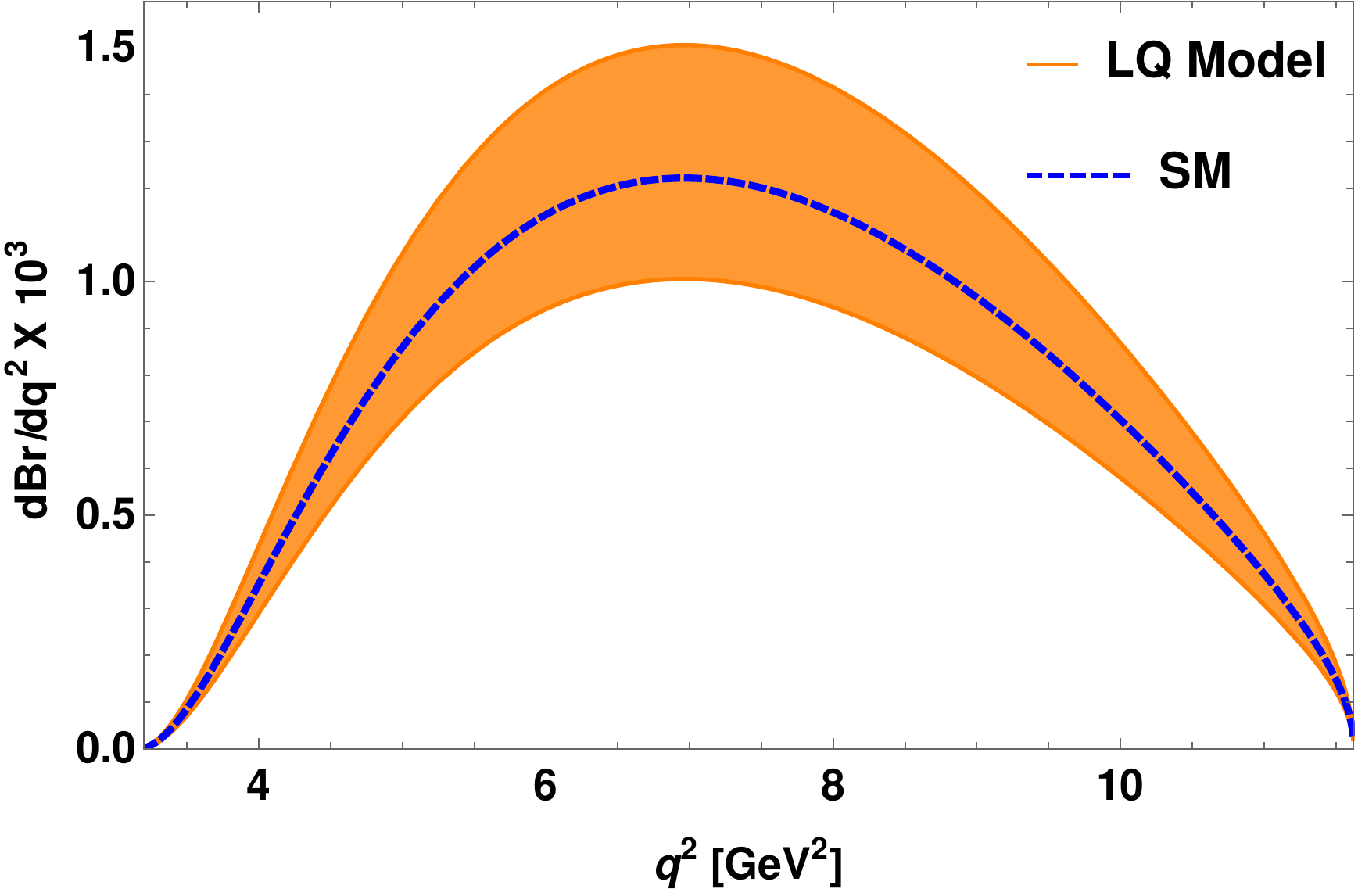}
\caption{The variation of  branching ratio of $B \to D \mu \bar{\nu_l}$ process (left panel) and $B \to D \tau \bar{\nu_l}$ process (right panel) with respect to $q^2$ in the leptoquark model. Here dashed lines are for SM and bands represent  leptoquark model.}
\end{figure}
\begin{figure}[h]
\centering
\includegraphics[width=7.5 cm,height=5.5cm]{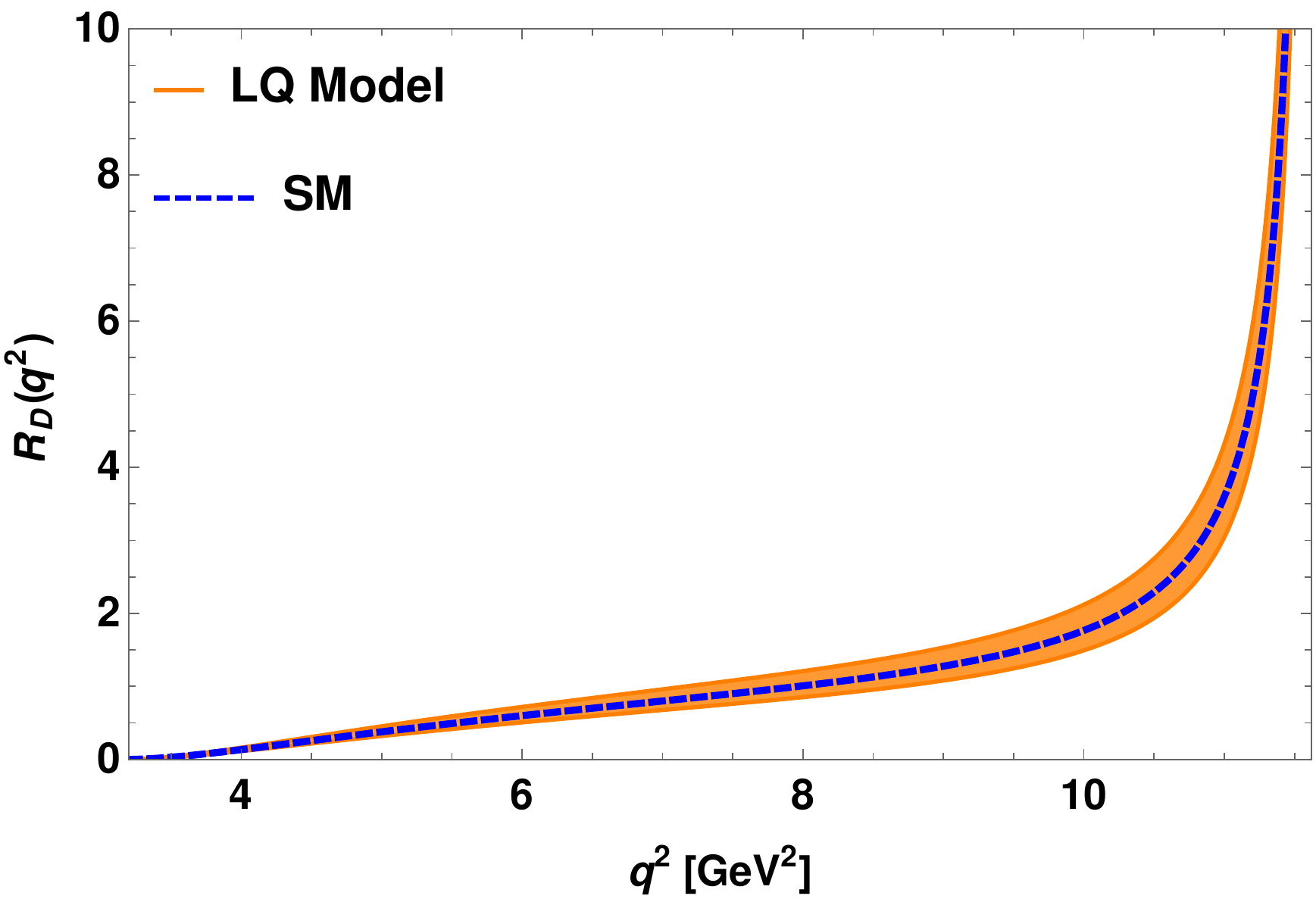}
\quad
\includegraphics[width=7.5 cm,height=5.5cm]{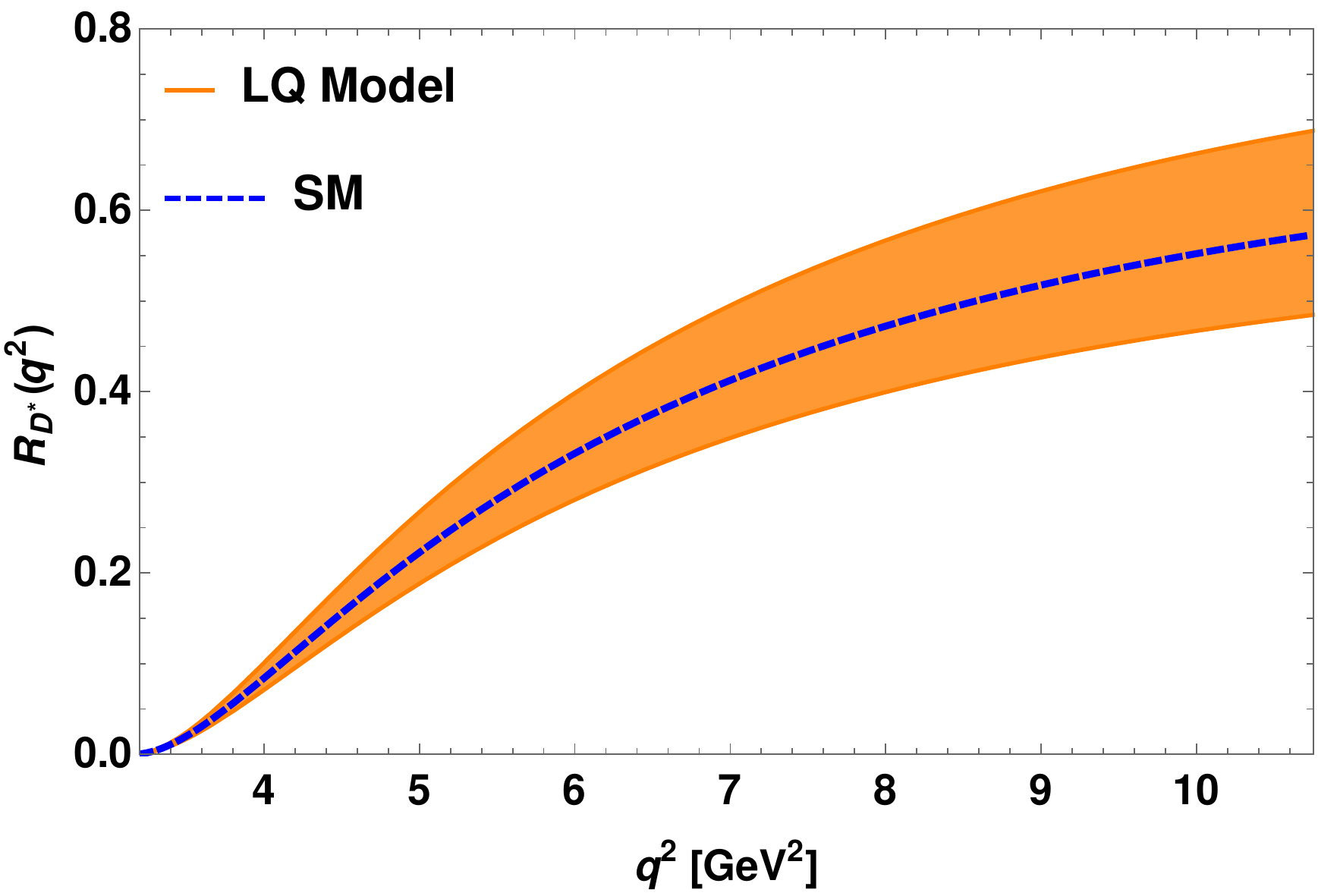}
\caption{The  $q^2$ variation of lepton non-universality $R_D (q^2)$ (left panel) and $R_{D^*}(q^2)$ (right panel) in  leptoquark model.}
\end{figure}
\begin{figure}[h]
\centering
\includegraphics[width=7.5 cm,height=5.5cm]{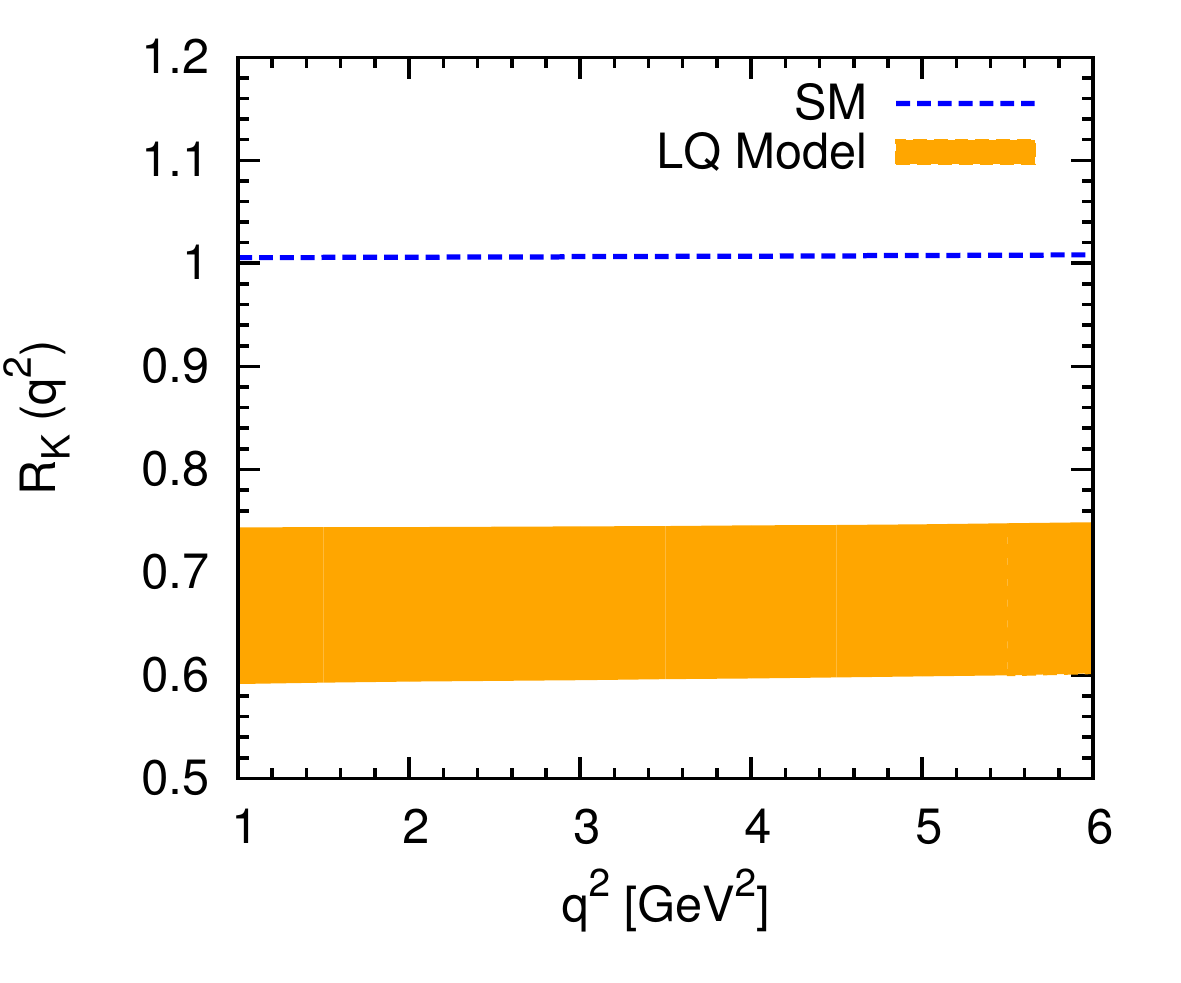}
\quad
\includegraphics[width=7.5 cm,height=5.5cm]{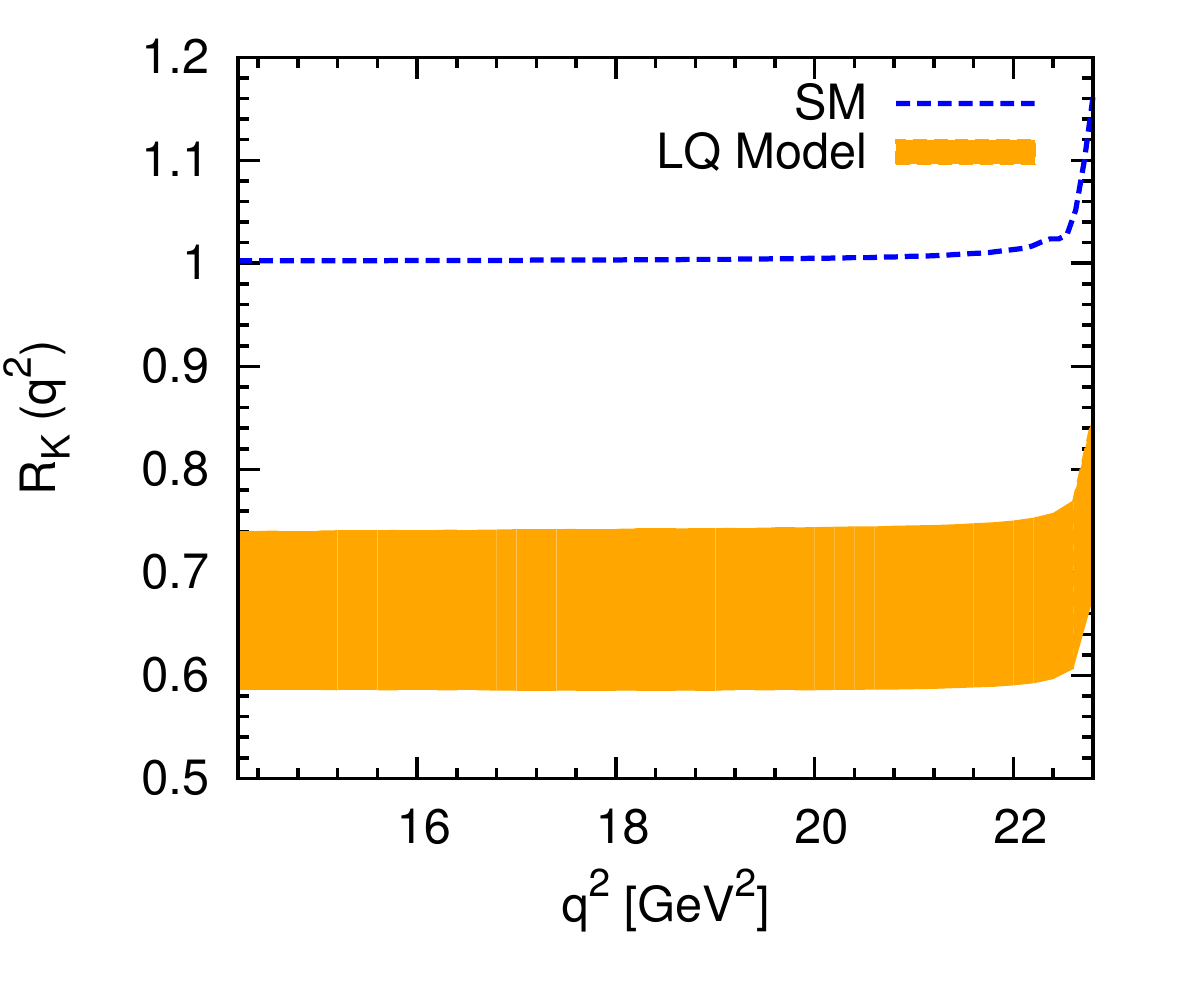}
\caption{ $R_K^{\mu e} (q^2)$ in low $q^2$( left panel)and  high $q^2$ (right panel)  in the leptoquark model.}
\end{figure}
\begin{figure}[h]
\centering
\includegraphics[width=7.0 cm,height=5.5cm]{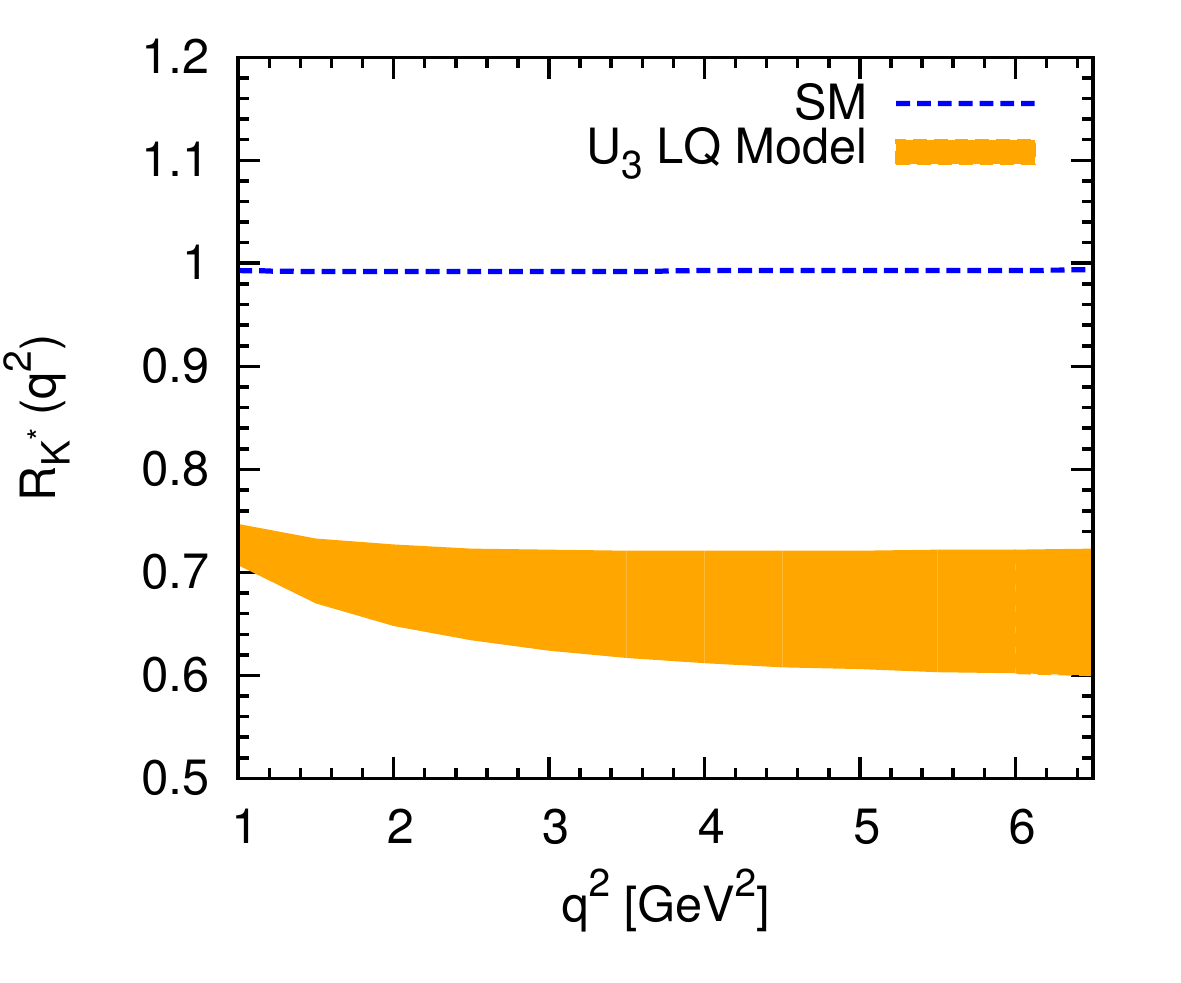}
\quad
\includegraphics[width=7.0 cm,height=5.5cm]{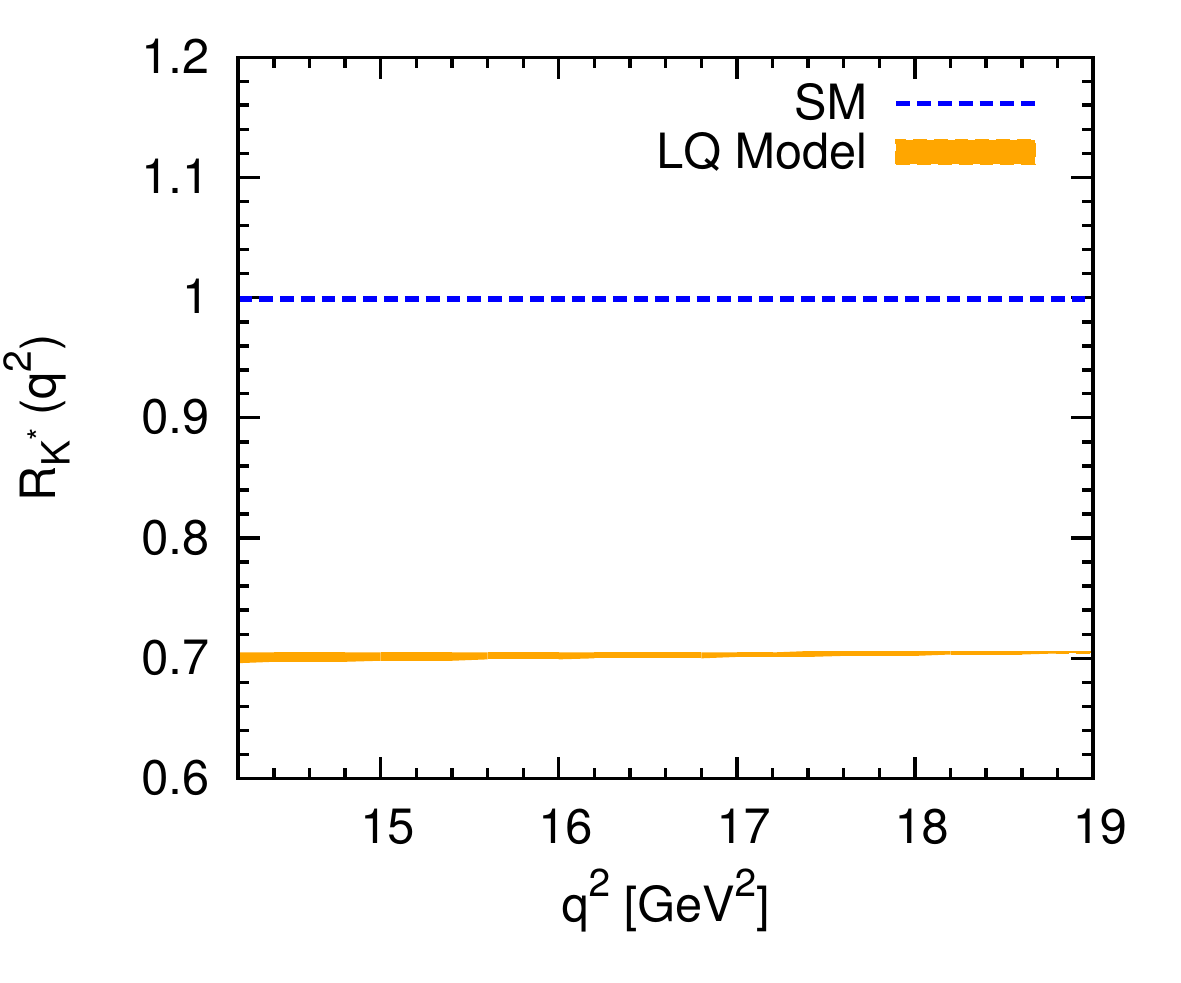}
\caption{ $R_{K^*}^{\mu e} (q^2)$ in low $q^2$ (left panel) and high $q^2$ ( right panel) in the leptoquark model.}
\end{figure}

The deviations in the observables $R_D$ and $R_{D^*}$, in comparison to that of the SM predictions, have been reported sometime ago both by Belle and BaBar. With the announcement of new result from the LHCb the situation has not changed anyhow and as a matter of fact the combined deviation from all three experiments is still more than 3$\sigma$ away from the SM expectation. At the same time we have also noticed the lepton nonuniversality in the form of $R_K$, where both first and second generation of leptons are involved. Past studies in the literature have
indicated many scenarios for the possible reason behind such discrepancies. In a model independent analysis it was shown that new physics in the form vector type of couplings could be a possible candidate option for such a discrepancy.  In this report we have considered the vector leptoquark model to explain the discrepancies obtained in the so-called $R_D$ and $R_{D^*}$ problems and also the lepton nonuniversality observable $R_K$. It is interesting to note that we can simultaneously explain both these anomalies, one in the tree level decay and another one in the loop suppressed process, using the vector leptoquarks. Therefore, using the scenario of vector leptoquarks, as the possible new physics candidature, one can attempt to understand the current  discrepancies in the beauty sector related to lepton non-universality and we hope more refined measurements will resolve these puzzles in the next few years (using additional data obtained from LHCb and Belle II) 
or else will give some smoking gun signal for physics beyond the SM.\\

{\bf Acknowledgments}\\

The authors would like to thank Science and Engineering Research Board (SERB),
Government of India for financial support through grant No. SB/S2/HEP-017/2013.\\

{\bf References}\\

\end{document}